\documentclass[conference]{IEEEtran}
\IEEEoverridecommandlockouts

\makeatletter
\def\markboth#1#2{\def\leftmark{\@IEEEcompsoconly{\sffamily}\MakeUppercase{\protect#1}}%
\def\rightmark{\@IEEEcompsoconly{\sffamily}\MakeUppercase{\protect#2}}}
\makeatother

\usepackage[english]{babel}
\usepackage{color}
\usepackage{lipsum}
\usepackage{caption}
\usepackage{cite}
\usepackage[pdftex]{graphicx}
\usepackage{subcaption}
\usepackage{amsmath}
\usepackage{amsfonts}
\usepackage{amssymb}
\usepackage{amsthm}
\usepackage{array}
\usepackage{verbatim}
\usepackage{listings}
\usepackage{algorithm}
\usepackage{algpseudocode}
\usepackage{url}
\usepackage{enumerate}
\usepackage{multirow}

\usepackage{epsfig}
\usepackage{epstopdf}
\usepackage{multicol}
\usepackage[font=footnotesize]{caption}

\renewcommand{\arraystretch}{2}

\newcommand{\bi}{\begin{itemize}}
\newcommand{\ei}{\end{itemize}}
\newcommand{\be}{\begin{equation}}
\newcommand{\ee}{\end{equation}}

\def\beq{\begin{equation}}
\def\eeq{\end{equation}}
\def\beqa{\begin{eqnarray}}
\def\eeqa{\end{eqnarray}}
\def\beqan{\begin{eqnarray*}}
\def\eeqan{\end{eqnarray*}}

\usepackage{threeparttable}
\usepackage[table,xcdraw]{xcolor}
\usepackage{tabularx}
   \usepackage{multirow}
   \usepackage{booktabs}
   \usepackage{graphicx}
   \usepackage{array, blindtext}
   \usepackage{wrapfig}
\usepackage{pdfpages}
\usepackage{hyperref}

\title{Comparative Analysis of Initial Access Techniques \\ in 5G mmWave Cellular Networks}
\author{{{\bf Marco Giordani}$^\dagger$, {\bf Marco Mezzavilla}$^\diamond$, {\bf C. Nicolas Barati}$^\diamond$, {\bf Sundeep Rangan}$^\diamond$, {\bf Michele Zorzi}$^\dagger$ }\\
$^\dagger$ University of Padova, Italy \quad $^\diamond$NYU Wireless, Brooklyn, NY, USA\\
emails: \small{$\{$\texttt{giordani}, \texttt{zorzi}$\}$\texttt{@dei.unipd.it}, $\{$\texttt{mezzavilla}, \texttt{srangan}, \texttt{nicolas.barati}$\}$\texttt{@nyu.edu}
}
}

\begin{document}
\maketitle

\begin{abstract}
The millimeter wave frequencies (roughly above $10$ GHz) offer the availability of massive bandwidth to greatly increase the capacity of fifth generation (5G) cellular wireless systems.  However, to overcome the high isotropic pathloss at these frequencies, highly directional transmissions will be required at both the base station (BS) and the mobile user equipment (UE)  to establish sufficient link budget in wide area networks.  This reliance on directionality has important implications for control layer procedures.  Initial access  in particular can be significantly delayed due to the need for the BS and  the UE to find the initial directions of transmission.  This paper provides a survey of several recently proposed techniques.  Detection probability and delay analysis is performed to compare  various techniques including exhaustive  and iterative search.  We show that the optimal strategy depends on the target SNR regime.
\end{abstract}


%
\IEEEpeerreviewmaketitle

\section{Introduction}
%
%
%
%
\IEEEPARstart{T}{he} fifth generation of cellular systems (5G) is positioned to address the demands and business contexts of 2020 and beyond. 
In order to face the continuing growth in demand from subscribers for a better mobile broadband experience, there is a need to push the performance to provide, for example \cite{nokia}: (i) much greater throughput, at least $1$ Gbps or more data rates, to support ultra-high definition video and virtual reality applications; (ii) much lower latency, less than $1$ ms, to support real time mobile control and Device-to-Device (D2D) applications and communications; (iii) ultra-high reliability and much higher connectivity; (iv) lower energy consumption, reduced by a factor of $1000$, to improve the battery life of connected devices.

In order to deal with these requirements, some key aspects have been identified to make this future network a reality. Since current micro-wave ($\mu$W) spectrum under $5$ GHz is  fragmented and crowded,  there has been significant interest in the millimeter wave (mmWave) bands above 10 GHz, where a vast amount of largely unused spectrum is available. On one hand, the enormous amount of available spectrum can support the higher data rates required in future mobile broadband access networks. Moreover, the physical size of antennas at mmWave frequencies is so small that it becomes practical to build very large antenna arrays (e.g., $\geq 32$ elements) to provide further gains from spatial isolation and multiplexing. On the other hand, to overcome the high isotropic pathloss of the mmWave frequencies, outdoor cellular links will need highly directional transmissions.  This complicates several control layer procedures, but is particularly challenging for initial access (IA).  Initial access is the procedure by which a mobile user equipment (UE) establishes a physical link connection with a base station (BS).  In current LTE systems, IA can be performed on omni-directional channels.  Beamforming or other directional transmissions can be performed after a physical link  is established.  However, in the mmWave range, the IA procedure must provide a mechanism by which the BS and UE can determine suitable initial directions of transmission.  This directional search can significantly delay the cell search and access procedure.

This paper provides a survey of recent directional IA techniques for mmWave cellular systems.  We compare various search schemes, including exhaustive search and an iterative scheme that successively narrows the search direction.  We study the performance in terms of misdetection probability and discovery delay, under some overhead constraints and as a function of the channel conditions.  Our results show that the optimal strategy depends on the target SNR regime and provide some guidance about the best scheme to use, according to the scenario.

The paper is organized as follows.  In Section \ref{sec:rel_work}, we review some of the most important contributions related to IA, while in Section \ref{sec:proc} we describe in detail the two initial access procedures we are going to compare. In Section \ref{sec:sim}  we evaluate through simulations some comparison metrics, such as misdetection probability and discovery delay, and finally, in Section \ref{sec:conclusions}, we summarize our major findings.

\section{Related work}
\label{sec:rel_work}

Papers  on IA are very recent, since research in this field is just at the dawn. Most literature refers to challenges that have been analyzed in the past at lower frequencies in ad hoc wireless network scenarios or, more recently, referred to the 60 GHz IEEE 802.11ad WLAN and WPAN scenarios. 

The initial access problem in mmWave cellular networks has been considered, for example, in \cite{pa1}, where the authors proposed an exhaustive method to sequentially scan the $360^\circ$ angular space. In \cite{pa2}, a directional cell discovery procedure is proposed,  where base stations periodically transmit synchronization signals, potentially in time-varying random directions, to scan the angular space. Also in \cite{pa3} initial access design options are compared, considering different scanning and signaling procedures, to evaluate access delay and system overhead; the analysis demonstrates significant benefits of low-resolution fully digital architectures in comparison to single stream analog beamforming. In \cite{bf_2}, random directional beamforming is considered for mmWave multi-user MISO downlink systems. By using asymptotic techniques, the performance of the algorithm is analyzed, based on a uniform random line-of-sight channel model suitable for highly directional mmWave radio propagation channels.

Paper \cite{bf_sd}  analyzes some low-complexity beamforming approaches for initial UE discovery in MIMO systems, showing that  users with a reasonable link margin can be quickly discovered by the proposed design with a smooth roll-off in performance as the link margin deteriorates.

Again \cite{pa4} presents a two-phase hierarchical procedure, while \cite{pa5} describes an energy-efficient link configuration mechanism,  by determining the optimal number of beams in a proposed two-stage beam training algorithm. In \cite{code_7}, in order to alleviate the exhaustive search delay issue, two types of adaptive beam training protocols are proposed. For fixed modulation, the proposed protocol allows for interactive beam training, stopping the search when a local maximum of the power angular spectrum is found that is sufficient to support the chosen modulation/coding scheme (approaches to prioritize certain directions determined from the propagation geometry, long-term statistics, etc. are also presented). For adaptive modulation, the proposed protocol uses iterative multi-level beam training concepts for fast link configuration that provide an exhaustive search with significantly lower complexity. 
In \cite{code_14}, a beamforming and cell search method is proposed for a mobile station with an antenna array to compensate the high attenuation in a mmWave OFDM-based cellular system. 

Also in \cite{varie_2}, exploiting a certain sparsity of mmWave channels, a low-complexity beam selection method for beamforming by low-cost analog beamformers is derived. It is shown that beam selection can be carried out without explicit channel estimation, using the notion of compressive sensing. 

With respect to this prior literature, our goal is to compare multiple IA procedures, in order to determine which is the preferred one, in terms of coverage probability and delay, to be implemented when considering a realistic dense, urban, multi-path scenario.

\section{Initial Access in 5G-mmWave networks}
\label{sec:proc}

We assume a slot structure similar to the one described in \cite{pa3}. 
Then, we assume that the PSS  is transmitted periodically once every $T_{per}$ seconds for a duration of $T_{sig}$ seconds, in each transmission. In this work, we will consider potentially different signal periods. A detailed description of each parameter that we will use in our simulation can be found in Section \ref{sec:sim}.
In this work, we consider two schemes, inspired by similar ideas in \cite{pa1} and \cite{pa4}, with some modifications of the transmitting and receiving scheme, as explained below.

\subsection{Exhaustive search}
Exhaustive search performs a brute-force sequential beam searching: the BS has a predefined codebook of $N$ directions (each  identified by a BF vector) that  covers the whole angular space. 

The goal  is to identify the best   TX-RX beam pair for each BS to connect with each UE. Therefore, the BS sends messages in those $N$ directions, in different slots, through narrow beams, while the UE configures its antenna array in order to directionally receive said messages. Upon the reception of a PSS, the user evaluates the SNR and, if it is above a fixed threshold, feeds it back to the BS through a $PSS_{RX}$ message. After having scanned the whole $360^\circ$ angular space, the BS determines the best beam to directionally reach the UE, on the basis of the highest received SNR (which corresponds to a certain direction).

As a design choice, the BS is equipped with $64$ $(8 \times 8)$ antennas and can steer beams in $N = 16$ directions, exploiting its maximum achievable BF gain. The UE has a set of combining vectors that also cover the whole angular space. In order to compare two different techniques, the UE can receive PSSs through $4$ wide beams (using a $2 \times 2$ antenna pattern) or through $8$ narrower beams (using all the $4 \times 4$ antenna elements it is equipped with).

\begin{table}[t!]
\centering
\begin{tabular}{*{3}{m{0.45\columnwidth}}}
\cline{1-2}
\textbf{Step 1 - DL:} BS transmits a PSS in sector $k=3$/16, UE receives in sector $1$.&\begin{center} {\includegraphics[width=.15\textwidth]{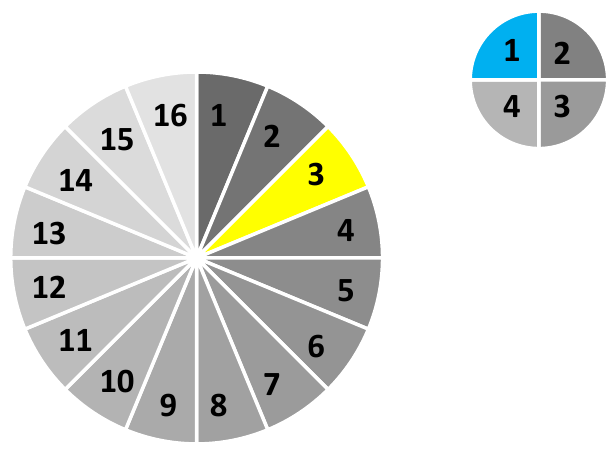} }\end{center} &\\
\cline{1-2}
\textbf{Step 2 - DL:} BS transmits a PSS in sector $k=3$/16, UE receives in sector $2$.&\begin{center} {\includegraphics[width=.15\textwidth]{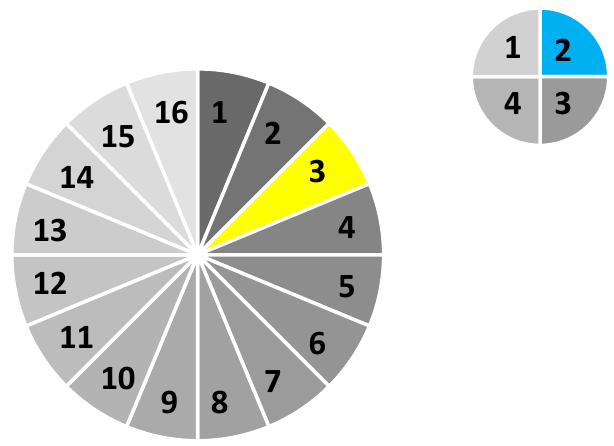} }\end{center} &\\
\cline{1-2}
\textbf{Step 3 - DL:} BS transmits a PSS in sector $k=3$/16, UE receives in sector $3$.&\begin{center} {\includegraphics[width=.15\textwidth]{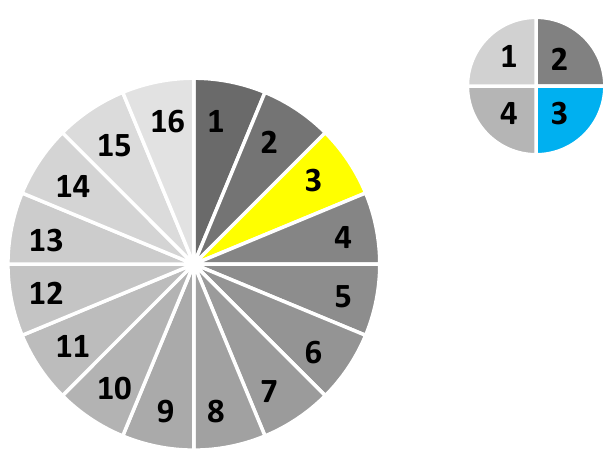} }\end{center} &\\
\cline{1-2}
\textbf{Step 4 - DL:} BS transmits a PSS in sector $k=3$/16, UE receives in sector $4$.&\begin{center} {\includegraphics[width=.15\textwidth]{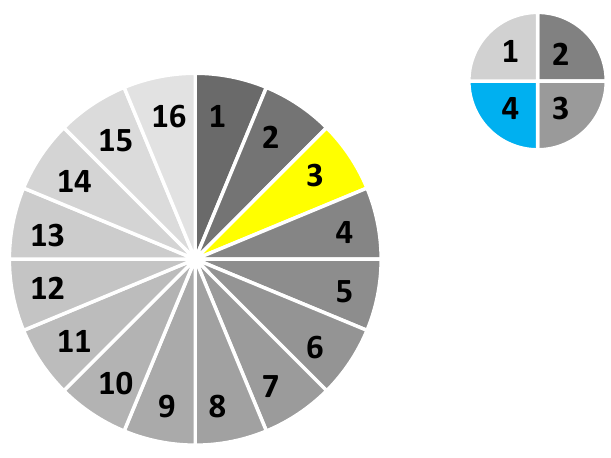} }\end{center} &\\
\cline{1-2}
\begin{tabular}{*{1}{m{0.95\columnwidth}}}
\vspace{0.6cm}
\textbf{Step 5 - Decision phase: }UE selects the sector in which it has perceived maximum SNR (if $>$ threshold), in this case sector $4$. This sector will be chosen to reach the BS with maximum performance. Then, the user selects a RNTI within a set of possibile signatures which will be used  for the whole procedure. This is its temporany ID to reach the BS. 
\vspace{0.6cm}
\end{tabular}\\
\cline{1-2}
\vspace{0.4cm}
\textbf{Step 6 - UL:} BS receives in the same sector $k=3$/16, UE transmits in its best sector (say $4$). UE transmits the information on its maximum perceived SNR, together with its RNTI. BS collects such information into a \textbf{BS table}. From now on, it knows that this user can be reached through sector $k=3$/16 with this SNR. \vspace{0.4cm}  &\begin{center} {\includegraphics[width=.15\textwidth]{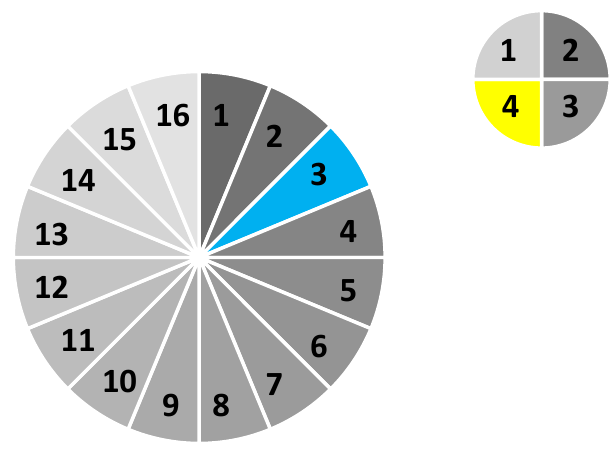} }\end{center} & \\
\cline{1-2}
\end{tabular}
\vspace{0.1cm}
\caption{Steps for exhaustive search, for a generic \emph{macroelement} $k=3$. All these steps will be hierarchically repeated for each  of the $16$ macroelements that the BS will send, in the corresponding angular directions.}
\label{tab:exh_step}
\end{table}

Let us consider a 4-direction reception (the 8-direction case is its natural extension). At the BS side, $4$ PSS messages are sent in the same direction, in $4$ consecutive DL slots, in order to match with one of the $4$ reception beams at the UE side. Then, in the UL slot, the user feeds back its $PSS_{RX}$, which is in turn collected at the BS side (in receiving mode). These $4$DL and $1$UL slots constitute an \emph{IA macroelement}. 
In Table \ref{tab:exh_step}, as an example, we just consider the macroelement $k=3$, referred to the BS steering direction $3/16$. However, the same steps will be hierarchically repeated for each one the $16$ macroelements that the BS will send.



After the UL slot in Step $6$, the UE returns to the reception mode, restarting the directional scan, looking for new PSSs. On the other hand, the BS configures its antenna elements in order to steer another macroelement block in the next direction $k + 1$ (Step 1 is cyclically performed again), until  the whole space is covered.

When all the $16$ directions have been scanned by the BS, both UE and BS select the best beam to reach each other, and start using it. In particular, the BS checks its BS table and finds, among the saved entries, the sector corresponding to the highest received SNR. This beam will be chosen to reach the UE.

\subsection{Iterative search}

Iterative search performs a two-stage scanning of the angular space. Again, a codebook is available, in order to send synchronization messages in deterministic directions.
 In the \emph{first phase}, the BS performs an exhaustive search through $4$ macro wide beams and, after having scanned  the whole $360^\circ$ space, determines its best beam, on the basis of the highest received SNR (similarly the UE finds the best direction to reach the BS). In the \emph{second phase}, the BS refines its search only in the previously identified wide beams.


In the first phase, the BS will send PSS messages in $N_1 = 4$ macro directions through $4$ wide beams, using $4$ antennas. In the second phase, instead, the BS will send the refining PSSs through $N_2 = 4$ narrow beams, through $64$ antennas. At the UE side, the user can receive PSSs through $4$ or $8$ beams, while in the second phase the UE will configure its antenna array to receive only in its best direction (previously identified).

Let us consider a $4$-direction reception (the $8$-direction case is its natural extension).  Similar to  the exhaustive procedure, in the first phase, an \emph{IA macroelement} is constituted of four DL slots (for  the four PSSs sent in the same direction, to match with one UE receiving beam) and one UL slot, reserved for the PSS$_{{RX}}$. Four macroelements are sent in the four $N_1$ directions, to cover the whole angular space. At the end, the BS determines the best macro direction $K$ to be refined, while the UE selects its best receiving beam.
In the second phase, the BS steers $N_2=4$ narrow beams to refine sector $K$: just one PSS per beam is sent, because now the UE no longer needs to scan in reception (in fact it will receive only through its best direction, determined during the first phase). 

Dwelling on the case where  UE receives in $4$ directions, in Table \ref{tab:it_step_ph1} we report the steps to be performed in the iterative \emph{first phase}, when considering, for example, macroelement $k=1$.
 \begin{table}[t!]
\centering
\begin{tabular}{*{3}{m{0.45\columnwidth}}}
\cline{1-2}
\textbf{Steps 1 to 4 - DL:} BS transmits a PSS in sector $k=1$/4, UE receives, in consecutive slots, in sectors $1$ to $4$, to cover the whole angular space.&\begin{center} {\includegraphics[width=.15\textwidth]{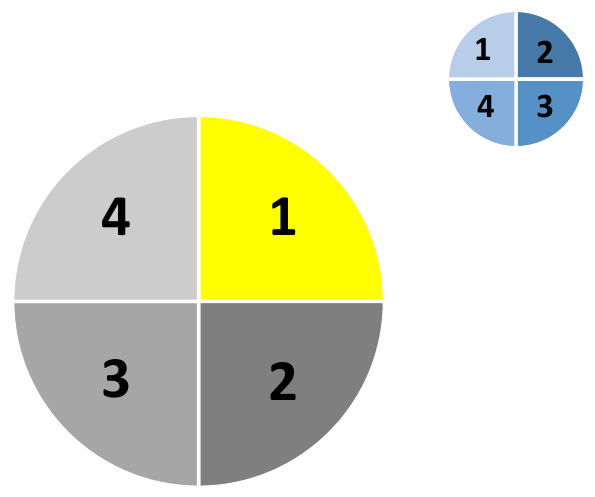} }\end{center} &\\
\cline{1-2}
\begin{tabular}{*{1}{m{0.95\columnwidth}}}
\vspace{0.6cm}
\textbf{Step 5 - Decision phase: }UE selects the sector in which it has perceived maximum SNR (if $>$ threshold), say sector $4$. This sector will be chosen to reach the BS with maximum performance. Then, the user selects a RNTI within a set of possibile signatures, which will be used for the whole procedure. This is its temporany ID to reach the BS. 
\vspace{0.6cm}
\end{tabular}\\
\cline{1-2}
\textbf{Step 6 - UL:} BS receives in the same sector $k=1$/4, UE transmits in its best sector (say $4$). UE transmits the information on its maximum perceived SNR, together with its RNTI. BS collects such information into a \textbf{BS table}. &\begin{center} {\includegraphics[width=.15\textwidth]{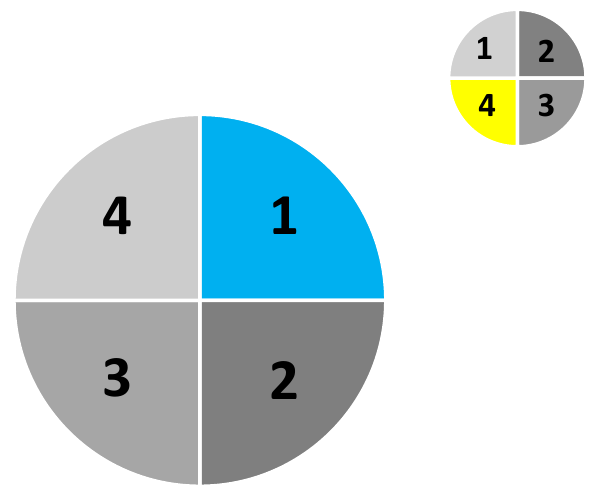} }\end{center} &\\
\cline{1-2}
\end{tabular}
\vspace{0.1cm}
 \caption{Steps for iterative search, in the first phase, for a generic \emph{macroelement} $k=1$. All these steps will be hierarchically repeated for each one of the $N_1=4$ wide sectors that the BS will send, in the corresponding angular directions.}
\label{tab:it_step_ph1}
\end{table}

    
    When all the $N_1=4$ macro directions have been scanned, UE selects its best receiving beam, to reach the BS, while the BS selects the best macro sector that will be refined in the second phase, whose iterative steps are shown in Table \ref{tab:it_step_ph2}.
    
      \begin{table}[!h]
\centering
\begin{tabular}{*{3}{m{0.45\columnwidth}}}
\cline{1-2}
\textbf{Step 1 - DL:} BS transmits one single refining PSS in small sector $k=1$/4, UE receives just through its best sector (in this case sector $4$) that has been previously identified in the first phase, without scanning the angular space. &\begin{center} {\includegraphics[width=.15\textwidth]{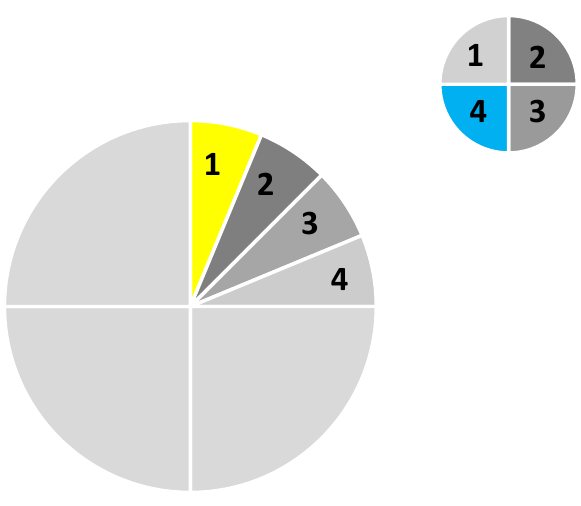} }\end{center}& \\
\cline{1-2}
\textbf{Step 2 - UL:} BS receives in the same sector $k=1$/4, UE transmits in its best sector $4$ the information on its maximum perceived SNR, together with its RNTI. BS collects again such info into its \textbf{BS table}.&\begin{center} {\includegraphics[width=.15\textwidth]{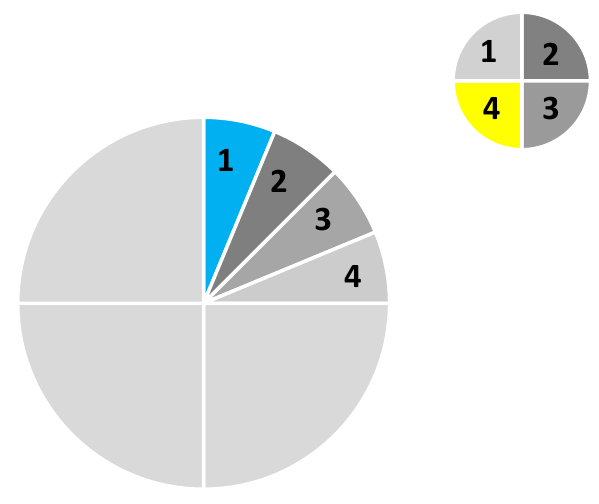} }\end{center}&\\
\cline{1-2}
\end{tabular}
\vspace{0.1cm}
    \caption{Steps for iterative search, in the second phase. All these steps will be hierarchically repeated for each one of the $N_2=4$ narrow sectors that the BS will send, in the corresponding refining directions.}
\label{tab:it_step_ph2}
\end{table}

After the UL slot, UE returns in reception mode and BS inspects the next refining direction, until  the whole  macro sector is scanned. Finally, the BS checks again its table and finds the sector (now a narrow beam) corresponding to the highest SNR saved. At this time, both BS and UE know how to directionally reach each other, in the best possible way.

\section{Simulation model and numerical results}
\label{sec:sim}

In the simulations, we will assume a static deployment, where no users are moving, so that no handover management or UE motion tracking is required. The parameters are based on realistic system design considerations and are summarized in Table \ref{tab:params}.

The channel model we have implemented is based on recent real-world measurements at $28$ GHz in New York City to provide a realistic assessment of mmWave micro and picocellular networks in a dense urban deployment. Statistical models are derived for key channel parameters including: (i) a distance-based pathloss, which models line-of-sight (LOS), non-line-of-sight (NLOS) and outage conditions; (ii) spatial clusters, described by central azimuth and elevation angles, fractions of power  and angular  beamspreads, (iii) a small-scale fading model, where each of the path clusters is synthesized with a large number of \emph{subpaths}, each one having its own peculiarities on horizontal and vertical angles (generated around the cluster central angles). Further details of the channel model and its parameters can be found in \cite{mustafa}.

A typical 5G cell is envisioned to have a radius of around $100 \div 200$ m, according to the channel characteristics. However, we will see that, referring to the channel described in \cite{mustafa}, much smaller cell radii should be considered, in order to provide good performance to most users.

In this work, we just refer to analog beamforming, that is implemented through Uniform Planar Array (UPA). A set of two dimensional antenna arrays is used at both BS and UE. The array can be comprised of $8 \times 8$, $4 \times 4$ or $2 \times 2$  elements. The spacing of the elements is set to $\lambda/2$, where $\lambda$ is the wavelength. These antenna patterns were chosen following the results in \cite{pa2} and can be shown to offer excellent system capacity for small cell urban deployment, together with easy packageability (for instance, at $28$ GHz, a $4 \times 4$ array will have a size of roughly $1.5$ cm $ \times 1.5$ cm).

Referring to the IA procedures, we will consider an SNR thresholds $\tau= -5$ dB. If the SNR is below $\tau$, it is assumed that the UE does not receive any PSS signal. Reduction of $\tau$ permits to find more users, at the cost of designing more complex (and expensive) receiving schemes, able to detect more corrupted messages.
Each signal has a minimum duration $T_{sig}=10\:\mu$s, that is sufficiently small for the channel to be coherent even at the very high frequencies used for mmWave communication \cite{pa3}. Moreover, beam switching time from one sector to another takes around $100$ ns, which is much less than $10 \: \mu$s and so can be neglected \cite{bf_1}.
Simulations are conducted increasing the distance of the UE from the BS, placed at coordinates $(0, 0)$ m,  in $10$ m increments. At each iteration, the user is deployed within an annulus having outer radius $R_1$ and inner radius $R_2$, with $R_2 < R_1$, according to a uniform distribution. In order to make reliable measurements through a Montecarlo estimation, each simulation is independently repeated  $10^6$ times.

\renewcommand{\arraystretch}{1}
\begin{table}[!t]
\centering
\begin{tabular}{*{3}{m{0.27\columnwidth}}}
\toprule
\textbf{Parameter} & \textbf{Value} & \textbf{Description}\\
\toprule
\rowcolor[HTML]{EFEFEF}  BW & $1$ GHz & Total system bandwidth\\
\midrule
DL - UL $P_{\text{TX}}$ & $30$ dBm & Transmission power \\
\midrule
\rowcolor[HTML]{EFEFEF}  NF  & $5$ dB & Noise figure \\ 
\midrule
$f_c$ & $28$ GHz & Carrier frequency \\
\midrule
\rowcolor[HTML]{EFEFEF}  $R$ & $0 \div 100 $ or $200$ m & Cell radius \\
\midrule
$\tau$ & $ -5$ dB &  SNR threshold \\
\midrule
\rowcolor[HTML]{EFEFEF}  BS antenna & $8 \times 8$  &  \\
\midrule
UE antenna & $4 \times 4$ or $2 \times 2$  & \\
\midrule
\rowcolor[HTML]{EFEFEF} BS position & $(0,0)$ m &  \\
\midrule
UE position & varied & Uniform in annulus \\
\midrule
\rowcolor[HTML]{EFEFEF}  UE speed & $0$ m/s & No mobility \\
\midrule
BF & analog & Beamforming architecture \\
\midrule
\rowcolor[HTML]{EFEFEF}  min. $T_{sig}$ & $10 \, \: \mu s$& min. signal duration \\
\midrule
$\phi_{ov}$ & $5\%$ & Overhead\\
\midrule
\rowcolor[HTML]{EFEFEF}$T_{per}$ & varied according to $\phi_{ov}$ & Period between transmissions \\
\midrule
Propagation loss model & LOS, NLOS, outage & According to \cite{mustafa} \\
\bottomrule
\end{tabular}
\caption{Simulation parameters for IA procedures.}
\label{tab:params}
\end{table}

In our study, we evaluate the performance in terms of discovery delay and misdetection probability. Discovery delay is the time a BS needs to identify all users in its coverage range, and can be computed as explained in subsection \ref{sec:delay}. 
The \emph{misdetection probability} (PMD)  is the probability that a UE within the cell is not detected by the BS, perceiving SNR below threshold. By making a large number of independent experiments, we keep a record of the number of times the UE's perceived SNR is below the threshold, to determine the corresponding PMD statistics.
We will consider different antenna array configurations, according to the reception mode the UE is programmed to use (either 4 or 8 receiving beams).


\subsection{Required number of slots}
\label{sec:delay}
First of all, we compare the different IA techniques from the discovery delay point of view. Considering a signal duration $T_{sig}=10\:\mu$s and a target \emph{overhead} of  $\phi_{ov}=5\%$, the time between two consecutive slot transmissions must be at least $T_{per} = T_{sig} / \phi_{ov} = 200\:\mu $s. Given that an IA procedure requires $N_s$ slots to be sent, the  delay in Table \ref{tab:req_slots} can be computed as $N_s \cdot T_{sig} / \phi_{ov} $.

\begin{table}[!h]
\centering
\begin{tabular}{c c c c c}
\toprule
\textbf{Procedure} & \textbf{TX antennas} & \textbf{RX antennas} & \textbf{$N_s$} &  \textbf{ Delay}   \\
\toprule
Exh. $64 \times 4$ & $64$& $4$  & $80$ & $16$ ms \\
\midrule
\rowcolor[HTML]{EFEFEF} Exh. $64 \times 16$ & $64$ & $16$ & $144$ & $28.8$ ms \\
\midrule
It. $64 \times 4$ & \begin{tabular}{@{}c @{}} $4$ in $1^{\text{st}}$ phase \\  $64$ in $2^{\text{nd}}$ phase  \end{tabular} &  $4$ & $28$ & $5.6$ ms\\
\midrule
\rowcolor[HTML]{EFEFEF} It. $64 \times 16$ & \begin{tabular}{@{}c @{}} $4$ in $1^{\text{st}}$ phase \\  $64$ in $2^{\text{nd}}$ phase  \end{tabular} &  $16$ & $44$ & $8.8$ ms\\
\bottomrule
\end{tabular}
\caption{Discovery delay for different IA techniques.}
\label{tab:req_slots}
\end{table}


We clearly see that iterative techniques outperform the exhaustive ones in terms of discovery delay. In fact, there is no need to scan the whole $360^\circ$ angular space to find the UE, since it is sufficient to just refine a  macro sector. 

\subsection{Misdetection probability}
\begin{figure}[t!]
\centering
 \includegraphics[trim= 0cm 0cm 0cm 0cm , clip=true, width=0.9 \columnwidth]{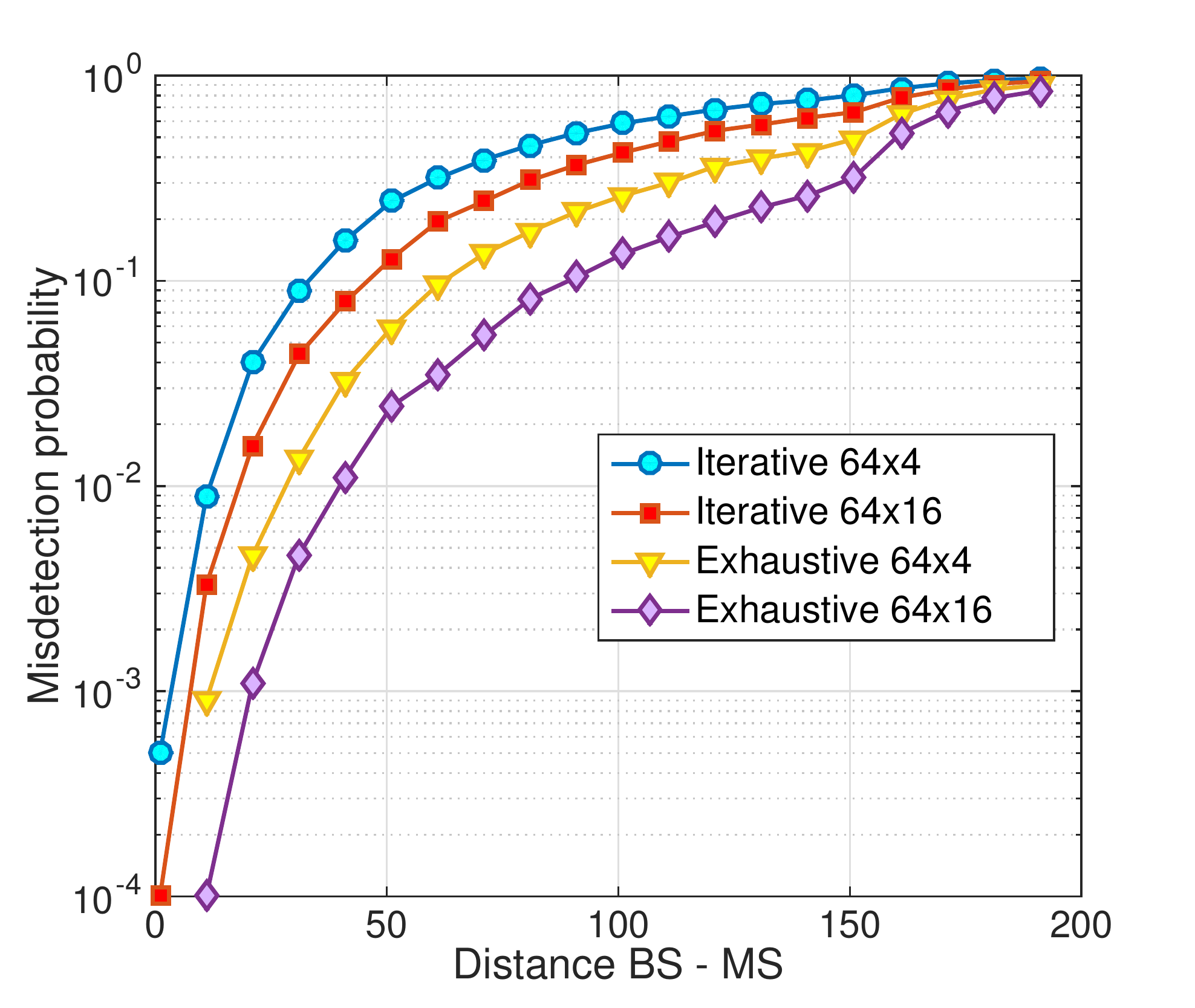}
 \caption{PMD for exhaustive and iterative techniques, when UE receives in  $4$ or $8$ directions, vs. the  BS-UE distance. SNR threshold $\tau=-5$ dB.}
 \label{fig:MDP_CS}
\end{figure}

Figure \ref{fig:MDP_CS} plots the misdetection probability of both exhaustive and iterative IA techniques, when UE receives in either $4$ or $8$ directions, varying the distance between BS and UE, from $0$ to $200$ meters. Threshold is  $\tau=-5$ dB.

The lower the number of antennas at both UE and BS (scanning through fewer directions in reception or transmitting through wider beams), the lower the BF gain, the lower the SNR perceived by the UE and the higher the probability that this SNR is below the threshold. For this reason, since iterative search makes use of a small antenna array in the first phase (just $4$ elements), it presents a higher misdetection probability, at the same distance, with respect to exhaustive searches.
However, in the range $0 \div 30$ meters, almost all algorithms present acceptable misdetection probability levels. At these distances, LOS condition is very likely met, and pathloss and  PMD are sufficiently small even when exploiting iterative searches. Therefore, when considering cells having very small radius, it may not be desirable to implement exhaustive procedures, which are affected by higher discovery delays.
In the range $100 \div 200$ meters, almost all algorithms present  instead unacceptable misdetection probability values, showing that better IA procedures need to be designed for cells of this size to be able to correctly operate.

Finally, we highlight the change of slope in the $64\times 16$ line, of Figure \ref{fig:MDP_CS}, around $150$ m. In fact, at this distance, the \emph{outage} pathloss condition is very likely to occur, with a  consequent considerable reduction of perceived SNR even for exhaustive scenarios where high BF is achieved, which explains the sudden increase of PMD.

\subsection{Trade-off between delay and PMD (total delay)}
\begin{figure}[t!]
\centering
 \includegraphics[trim= 0cm 0cm 0cm 0cm , clip=true, width=0.9 \columnwidth]{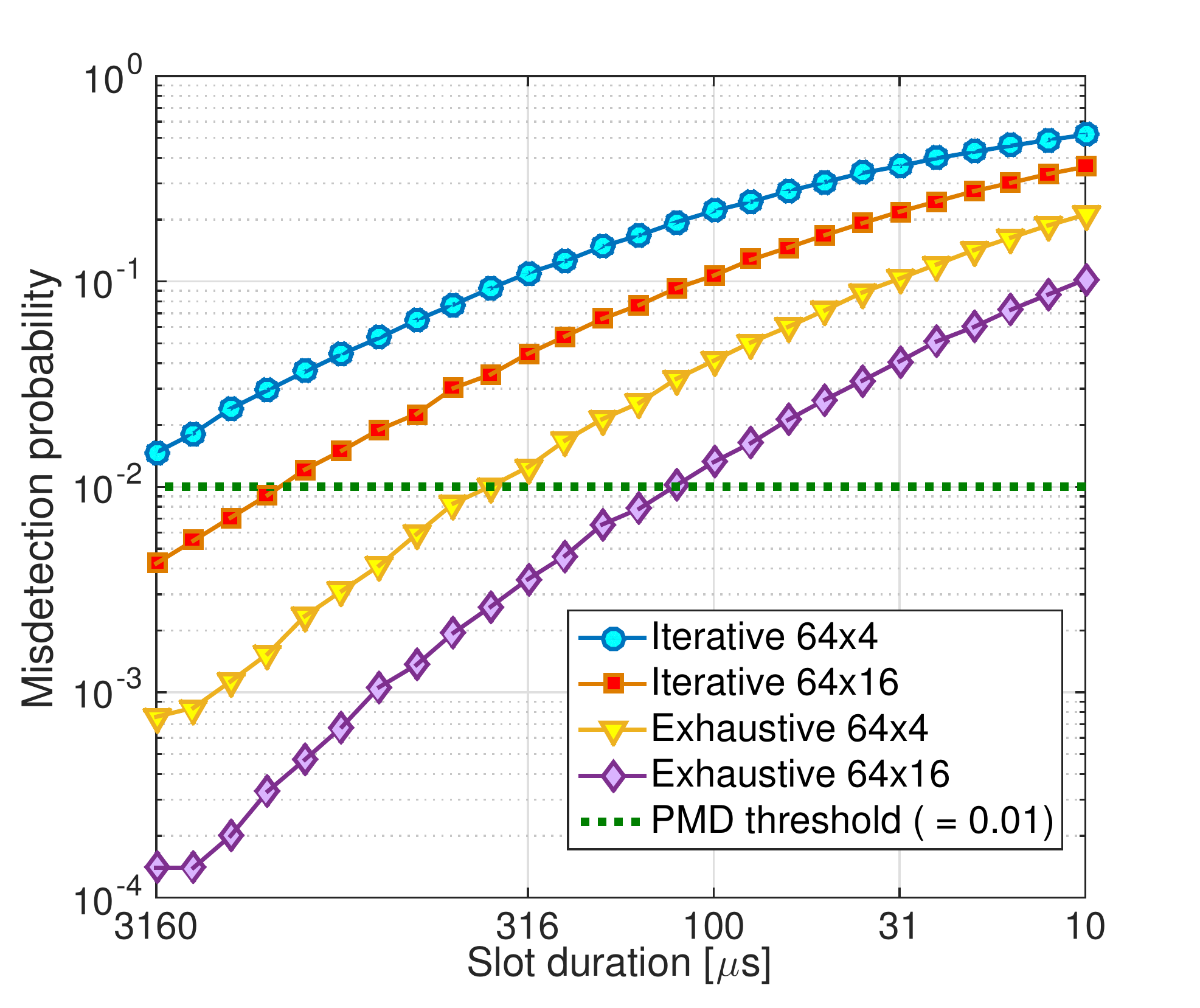}
 \caption{PMD for exhaustive and iterative searches, vs. signal duration $T_{sig}$. BS-UE distance $d=95$ m.}
 \label{fig:slot_dur}
\end{figure}

Considering now a worst-case scenario, we want to \emph{constrain} the misdetection probability to be  below a certain threshold (i.e., $< 0.01$) for most users in the cell. If the cell radius is around $100$ m, we want that PMD$ < 0.01$ \emph{even for edge users}, at $95$ meters from the BS.
In order to decrease the PMD, we can increase the signal duration: if $T_{sig}$ is increased, the BS transmits its PSS for a longer time in the same sector, so that UEs belonging to that sector can accumulate a higher amount of energy. This results in an increased SNR  and  in a reduced PMD. Increasing the signal duration is simulated by correspondingly lowering  $\tau$. In such a way, if the signal duration is doubled, the  SNR also doubles and this is equal to reducing $\tau$ by $3$ dB.
From the results in Figure \ref{fig:slot_dur}, we can compute the minimum signal length to meet the PMD requirements for edge users.

In Table \ref{tab:time}, we  compute the \emph{total delay} that takes care of both the required number of slots to perform each IA technique and the PMD specifications. According to the specific value of $T_{sig}$ that has been found, the corresponding value of $T_{per}$ should be chosen, to maintain a constant overhead of $5\%$.

\begin{table}[!h]
\centering
\begin{tabular}{*{4}{m{0.2\columnwidth}}}
\toprule
\textbf{Procedure} & \textbf{$N_s$} &  \textbf{min. $T_{sig}$} &  \textbf{Total delay $N_s \cdot T_{sig} / \phi_{ov} $}   \\
\toprule
\begin{tabular}{@{}c@{}}Exh. $64\times 4$   \end{tabular}  & $80$ & $400\: \mu$s & $640$ ms  \\
\midrule
\rowcolor[HTML]{EFEFEF} \begin{tabular}{@{}c@{}} \rowcolor[HTML]{EFEFEF} Exh. $64\times 16 $ \end{tabular} & $144$ & $125 \:\mu$s &$360$ ms \\
\midrule
\begin{tabular}{@{}c@{}}It. $64\times 4$   \end{tabular} & $28$ & $>3160\: \mu$s & $> 1760$ ms \\
\midrule
\rowcolor[HTML]{EFEFEF} \begin{tabular}{@{}c@{}} \rowcolor[HTML]{EFEFEF} It. $64\times 16$  \end{tabular}  & $44$  & $1580 \:\mu$s & $\simeq 1390$ ms\\
\bottomrule
\end{tabular}
\caption{Total delay, to guarantee PMD $< 0.01$ for edge users.}
\label{tab:time}
\end{table}

We can see that iterative techniques require very long signals, to overcome the low BF gain and collect enough energy to perceive a sufficiently high SNR. On the other hand, exhaustive searches can have shorter slots.

At least for this kind of scenarios, where the cell size is around $ 100$ m,  the total delay of iterative searches is  higher than that of the exhaustive scheme, which seems therefore to be preferable (if we want to guarantee good coverage probability even for {edge users}). Otherwise, if we want to grant a good PMD just for \emph{closer users}, iterative techniques will lead to a sufficiently high SNR even exploiting shorter slots.

Moreover, we highlight that the{ iterative technique is not recommended when dealing with very dense networks}.  
When a UE is detected in a certain macro direction, after the first phase, this sector is consequently refined. 
If multiple users are found in \emph{different} macro directions, it is necessary to refine all of them, one at a time. This way, discovery can take longer than with exhaustive techniques, and even with a worse misdetection probability, due to low first phase BF gain. It is advisable to implement such an iterative  procedure only when the users' arrival rate is low and when the probability that multiple UEs are under the coverage area of different macro sectors is negligible.

\section{Conclusions and future works}
\label{sec:conclusions}

In this work, we have studied, analyzed and compared possible implementations of initial access techniques for upcoming 5G millimeter-wave cellular networks. The main idea is that, in order to overcome the bad mmWave channel propagation conditions, directionality should be realized also in the initial synchronization-access phase. Our analysis has indeed demonstrated the following key findings:

\begin{itemize}
\item There exists a trade-off between IA delay and misdetection probability: on the one hand, iterative techniques require fewer slots to perform the angular search, with respect to exhaustive algorithms; on the other hand, they make use of a small antenna array in the first phase and present higher misdetection probability levels.

\item When wanting to grant a minimum coverage level also to edge users (say around $100$ meters from the BS) and when considering a  dense, urban, multipath mmWave channel, exhaustive procedures are preferred, since they achieve a smaller total delay, in comparison to other IA techniques.

\end{itemize}

As part of our future work, digital (or hybrid) beamforming  architectures will be studied to steer multiple narrow beams in multiple directions at the same time, in order to reconcile the low discovery delay goal  with low misdetection probability.
Furthermore, IA procedures where HetNets and Context-Information are exploited deserve a deeper investigation.


\section*{Acknowledgement}

Part of this work has been performed in the framework of the H2020 project METIS-II co-funded by the EU. The authors would like to acknowledge the contributions of their colleagues from METIS-II although the views expressed are those of the authors and do not necessarily represent the views of the METIS-II project.

\bibliographystyle{IEEEtran}
\bibliography{bibliography.bib}

\end{document}